\documentclass{ws-procs9x6}

\begin{document}

\title{Study of Scalar Mesons and Related Radiative Decays}

\author{D. BLACK$^a$, M. HARADA$^b$ AND J. SCHECHTER$^c$\footnote{\uppercase{S}peaker}}

\address{(A) Theory Group, Jefferson Laboratory, \\
12000 Jefferson Avenue, \\ 
Newport News, Virginia, 23606, USA\\ 
E-mail: dblack@jlab.org}

\address{(B)Department of Physics, \\ 
Nagoya University, \\
Nagoya 464-8602, Japan \\
E-mail: harada@eken.phys.nagoya-u.ac.jp}  

\address{(C)Physics Department, \\
Syracuse University, \\
Syracuse, NY 13244-1130, USA \\
E-mail: schechte@phy.syr.edu}

\maketitle

\abstracts{
After a brief review of the puzzling light
scalar meson sector of QCD, a brief summary will be given of a 
paper concerning radiative decays involving the light scalars.
There, a simple vector meson dominance model is constructed in an initial
attempt to relate a large number of the radiative decays
involving a putative scalar nonet to each other. As an application
it is illustrated why $a_0(980)-f_0(980)$ mixing is not expected to
greatly alter the $f_0/a_0$ production ratio for radiative $\phi$
decays.}

\section{Introduction}

Why might the subject of light scalar mesons be of interest to physicists now
that QCD is known to be the correct theory of Strong Interactions and the burning issue
is to extend the Standard Model to higher energies?  
Simply put, another goal of Physics is to
produce results from Theory which can be compared with Experiment. At very large
energy scales, the asymptotic freedom of QCD guarantees that a controlled
perturbation expansion is a practical tool, once the relevant "low energy stuff"
is suitably parameterized. At very low energy scales, for example close to the
threshold of $\pi \pi$ scattering. the running QCD coupling constant is expected
to be large and perturbation theory is not expected to work. Fortunately, a
controlled expansion based on an effective theory
with the correct symmetry structure- Chiral Perturbation Theory\cite{GL}-
seems to work reasonably well. The new information about Strong Interactions which
this approach reveals is closely related to the spectrum and flavor "family"
properties of the lowest lying pseudoscalar meson multiplet and was, in fact,
essentially known before QCD.
 
\begin{figure}
\centerline{\epsfxsize=3.9in\epsfbox{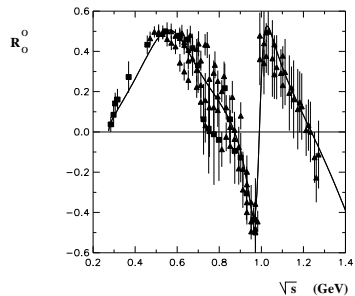}}
\caption{ Illustration of the real part of 
the pi pi scattering amplitude extracted from experimental data.}
\label{experiment}
\end{figure}

Clearly it is important to understand how far in energy above threshold 
the Chiral Perturbation Theory program will take us. To get a rough estimate
consider the experimental data for the real part of the $I=J=0$ $\pi \pi$
scattering amplitude, $R_0^0$ displayed in Fig.~\ref{experiment}. The chiral
perturbation series should essentially give a polynomial fit to this shape,
which up to about 1 GeV is crudely reminiscent of one cycle of a sine curve.

\begin{figure}
\vskip -2.0cm
\centerline{\epsfxsize=2.5in\epsfbox{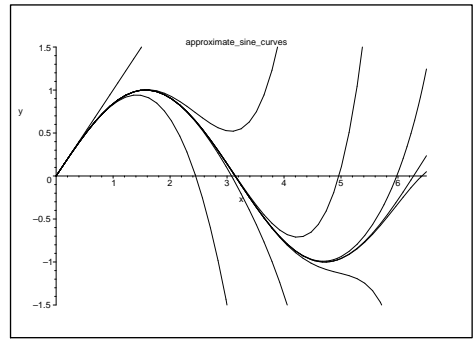}}
\vskip -2.5cm 
\caption{Polynomial approximations to one cycle of the sine curve.}
\label{polynomial}
\end{figure}

Now consider polynomial approximations to one cycle of the sine curve 
with various numbers of terms. These are illustrated in Fig.~\ref{polynomial}.
Note that each succesive term departs from the true sine curve right
after the preceding one. It is clear that something like eight terms
are required for a decent fit. This would correspond to seven loop order
of chiral perturbation theory and seems presently impractical.  

\section{Need for light scalar mesons}

Thus an alternative approach is indicated for going beyond threshold
of pi pi scattering up to about 1 GeV. The data itself suggests
the presence of s-wave resonances, the lowest of which is denoted the
"sigma". Physically, one then expects the practical range of
chiral perturbation theory to be up to about 450-500 MeV, just before the 
location of this lowest resonance.
In the last few years there have been studies
\cite{recentwork} by many authors which advance this picture. All
of them are "model dependent" but this is probably inevitable for
the strongly coupled regime of QCD.  For example \cite{HSS}, in a framework
where the amplitude is  computed from a non linear chiral Lagrangian
containing explicit scalars as well as vectors and pseudoscalars, the fit shown
in Fig.~\ref{experiment} emerges as a sum
of four pieces: i. the current algebra
"contact" term, ii. the $\rho$ exchange diagram iii. a non Breit Wigner
 $\sigma(560)$ pole
diagram and exchange, iv. an $f_0(980)$ pole in the background produced  
 by the other three. It is not just a simple sum of Born graphs but includes
the approximate unitarization features of the non Breit Wigner shape of the sigma 
and a Ramsauer Townsend mechanism which reverses the sign of the $f_0(980)$. 
Also note that i. and ii. provide very substantial background to the sigma
pole, partially explaining why the sigma does not "jump right out" of
various experimental studies. Qualitative agreement with this approach is obtained by
K-matrix unitarization of the two flavor linear sigma model \cite {ASII}
and three flavor linear sigma model \cite{BFMNS} amplitudes.

Workers on scalar mesons entertain the hope that, after the revelations
about the vacuum structure of QCD confirmed by the broken chiral symmetric
treatment of the pseudoscalars, an understanding of the next layer 
of the "strong interaction onion"
will be provided by studying the light scalars. An initial question is
whether the light scalars belong to a flavor $SU(3)$ multiplet as the
underlying quark structure might suggest. Apart from the $\sigma(560)$,
the $f_0(980)$ and the isovector $a_0(980)$ are fairly well established. This leaves
a gap concerning the four strange- so called kappa- states. This question is
more controversial than that of the sigma state . In the unitarized 
non linear chiral       
Lagrangian framework one must thus consider $pi-K$ scattering.
 In this case the low energy amplitude is taken\cite{BFSS}
to correspond to the sum of a current algebra contact diagram,
vector $\rho$ and $K^*$ exchange diagrams and scalar $\sigma(560)$,
$f_0(980)$ and $\kappa(900)$ exchange diagrams. The situation in the
interesting $I=1/2$ s-wave channel turns out to be very analogous to
the $I=0$ channel of s-wave $\pi\pi$ scattering. Now a non Breit Wigner 
$\kappa$ is required to restore unitarity; it
plays the role of the $\sigma(560)$ in the $\pi\pi$ case. It was found that
a satisfactory description of the 1-1.5 GeV s-wave region is also
obtained by including the well known $K_0^*(1430)$ scalar
resonance, which plays the role of the $f_0(980)$ in the $\pi\pi$
calculation. As in the case of the sigma, the light kappa seems hidden by
background and does not jump right out of the initial analysis of the 
experimental data.

 Thus the nine states
 associated with the $\sigma(560)$, $\kappa(900)$,
$f_0(980)$ and $a_0(980)$ seem to be required in order to fit experiment   
in this chiral framework. What would their masses and coupling constants suggest
about their quark substructure if they were assumed to comprise an SU(3) nonet
\cite{putative}?
 Clearly the mass ordering of the various
states is inverted compared to the "ideal mixing"\cite{Okubo}
scenario which approximately holds for most meson nonets. This means that
a quark structure for the putative scalar nonet of the form  $N^b_a \sim q_a{\bar
q}^b$ is unlikely since the mass ordering just corresponds to counting the
number of heavier strange quarks. Then the nearly degenerate $f_0(980)$ and $a_0(980)$
which must have the structure $N^1_1 \pm N^2_2$ would be lightest rather
than heaviest. However the inverted
ordering will agree with this counting if we assume that the scalar mesons   
are schematically constructed as $N^b_a \sim T_a{\bar T}^b$ where $T_a
\sim \epsilon_{acd}{\bar q}^c{\bar q}^d$ is a "dual" quark (or
anti diquark). This interpretation is strengthened by consideration
\cite{putative} of the scalars' coupling constants to two
pseudoscalars. Those couplings depend on the value of a mixing angle, $\theta_s$
between $N^3_3$ and $(N^1_1-N^2_2)/{\sqrt 2})$. Fitting 
 the coupling constants
to the treatments of $\pi\pi$ and $K\pi$ scattering gives a mixing angle such that
 $\sigma \sim N^3_3 +$ "small"; $\sigma$(560) is thus
a predominantly non-strange particle in this picture. Furthermore   
the states $N^1_1 \pm N^2_2$ now would each predominantly contain two extra
 strange quarks and would be expected to be heaviest.
  Four quark pictures of various types have been sugggested as arising from
spin-spin interactions in the MIT bag model\cite{Jaffe}, 
unitarized quark models\cite{uqm} and
 meson-meson interaction models\cite{mmim}.

There seems to be another interesting twist to the story of the light scalars. 
 The success of the phenomenological quark model suggests
that there exists, in addition, a nonet of ``conventional"
p-wave $q{\bar q}$ scalars in
the energy region above 1 GeV. The experimental candidates for these states
are $a_0(1450)(I=1)$, $K_0^*(1430)(I=1/2)$ and for $I=0$,
$f_0(1370)$, $f_0(1500)$ and $f_0(1710)$. These are enough for a full
nonet plus a glueball. However it is puzzling that the strange
$K_0^*(1430)$ isn't noticeably heavier than the non strange $a_0(1450)$
and that they are not lighter than the corresponding spin 2 states.
These and another puzzle may be solved in a natural way\cite{BFS}
if the heavier p-wave scalar nonet mixes with a lighter $qq{\bar q}
{\bar q}$ nonet of the type mentioned above. The mixing mechanism makes  
essential use of the "bare" lighter nonet having an inverted mass ordering
while the heavier "bare" nonet has the normal ordering.
A rather rich structure involving
the light scalars seems to be emerging. At lower energies one may consider
as a first approximation, "integrating out" the heavier nonet and retaining
just the lighter one.

\section{Radiative decays involving light scalars}

In the last few years, a lot of experimental
activity\cite{ea} at the $e^+ e^-$ machines (Novosibirsk, DA$\Phi$NE
and Jefferson Lab) has resulted in definitive measurements of the
interesting reactions:

\begin{eqnarray}
&&
\phi(1020)\rightarrow f_0(980) +\gamma\rightarrow \pi^0\pi^0 +\gamma,
\\
&&
\phi(1020)\rightarrow a_0(980) +\gamma\rightarrow \pi^0\eta +\gamma.
\end{eqnarray}

These measurements have been awaited by theorists for a number of years as 
proposed tests \cite{scalartests}
of the nature of the $f_0(980)$ and $a_0(980)$ scalars. The theoretical models
used for these tests were based on the observation that the 
vector meson, $\phi(1020)$ mainly decays into
$K+{\bar K}$ so a virtual $K$ loop diagram
can reasonably be expected to dominate the decay mechanism. In this framework
Achasov \cite{Ach} has argued that the data are most consistent with a  
compact four quark structure for the $f_0$ and $a_0$ (as opposed to a two
quark structure or a loosely bound meson meson "molecule" structure).

This situation makes it interesting to study in detail the extension 
of the picture
to a full nonet (or two?) of scalar mesons as well as to further solidify
the technical analysis of the K- loop class of diagrams. In addition, there
is perhaps (depending on the exact masses and widths of the $a_0$ and $f_0$ mesons)
a problem in that the experimentally derived ratio $ \Gamma(\phi
\rightarrow f_0\gamma)/\Gamma(\phi \rightarrow a_0\gamma)$ is in the range
3-4 while theoretical estimates are mostly clustered around unity.
We are presently working on K-loop type models but decided to start
for ourselves with a much simpler preliminary picture. The goal of
this model \cite{BHS} is to try to correlate many different radiative processes
involving the members of a  full scalar nonet by using flavor symmetry.   
The model has the following features: 1. It is based on a
 chiral symmetric Lagrangian containing complete nonets of pseudoscalar,
vector as well as (the putative) scalar fields. 
2.Vector meson dominance for photon vertices is automatic in the formulation.
3. An effective flavor invariant SVV (scalar-vector-vector) vertex is
 postulated which has three relevant parameters. These are treated as the only
{\it a priori} unfixed parameters of the model.

\begin{figure}[htbp]
\centerline{\epsfxsize=3.9in\epsfbox{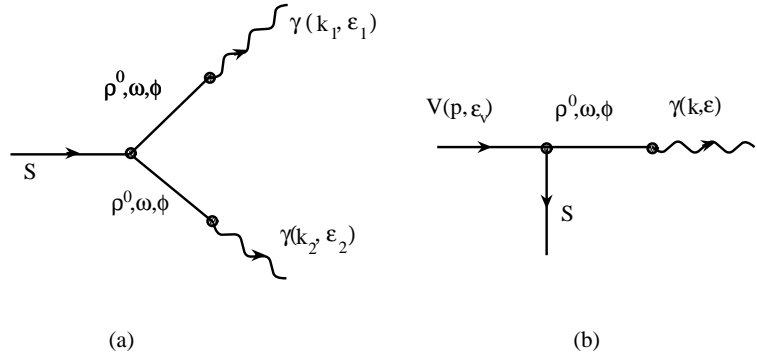}} 
\caption[]{%
Feynman diagrams for (a)~$S\rightarrow \gamma\gamma$ and
(b)~$V \rightarrow S \gamma$.
}\label{fig:diagrams}
\end{figure}

Our framework is that of a standard non-linear chiral Lagrangian   
containing, in addition to the pseudoscalar nonet matrix field $\phi$,
the vector meson nonet matrix $\rho_\mu$ and a scalar nonet matrix  
field denoted $N$.
Under chiral unitary transformations of the three light quarks;
$q_{\rm L,R} \rightarrow U_{\rm L,R} q_{\rm L,R}$, the chiral
matrix $U = \exp ( 2 i \phi/F_\pi)$,
where $F_\pi \simeq 0.131\,\mbox{GeV}$, transforms as
$U \rightarrow U_{\rm L} U  U_{\rm R}^\dagger$.
The convenient matrix
$K(U_{\rm L}, U_{\rm R}, \phi )$
is defined by the following transformation property of
$\xi$ ($U = \xi^2$):
$\xi \rightarrow U_{\rm L} \xi K^{\dag}
  = K \xi U_{\rm R}^{\dag}$,
and specifies the transformations of ``constituent-type'' objects.
The fields we need transform as
\begin{eqnarray}
&&
  N \rightarrow K N K^{\dag} \ ,
\nonumber\\
&&
  \rho_\mu \rightarrow K \rho_\mu K^{\dag}
  + \frac{i}{\tilde{g}} K \partial_\mu K^{\dag}
\ ,
\nonumber\\
&&
  F_{\mu\nu}(\rho) =
  \partial_\mu \rho_\nu - \partial_\nu \rho_\mu - i
  \tilde{g} \left[ \rho_\mu \,,\, \rho_\nu \right] 
  \rightarrow
  K F_{\mu\nu} K^{\dag}
\ ,
\label{transf}  
\end{eqnarray}
where the coupling constant $\tilde{g}$ is about $4.04$.
The strong trilinear
 scalar-vector-vector terms in the effective Lagrangian are:
\begin{eqnarray}
 &&
{ L}_{SVV} =  \beta_A \, 
\epsilon_{abc} \epsilon^{a'b'c'}
\left[ F_{\mu\nu}(\rho) \right]_{a'}^a
\left[ F_{\mu\nu}(\rho) \right]_{b'}^b N_{c'}^c
\nonumber\\
&&
 \quad
{}+
 \beta_B \, \mbox{Tr} \left[ N \right]
\mbox{Tr} \left[ F_{\mu\nu}(\rho) F_{\mu\nu}(\rho) \right]   
\nonumber\\
&&
 \quad
{}+
 \beta_C \, \mbox{Tr} \left[ N F_{\mu\nu}(\rho) \right]
\mbox{Tr} \left[ F_{\mu\nu}(\rho) \right]
\nonumber\\
&&
 \quad
{}+
 \beta_D \, \mbox{Tr} \left[ N \right]
\mbox{Tr} \left[ F_{\mu\nu}(\rho) \right]
\mbox{Tr} \left[ F_{\mu\nu}(\rho) \right]
\ .
\label{SVV}
\end{eqnarray}
Chiral invariance is evident from (\ref{transf}) and the four
flavor-invariants are needed for generality.  (A term
$\sim \mbox{Tr}( FFN )$ is linearly dependent on the four shown).
Actually the $\beta_D$ term will not contribute in our model so there
are only three relevant parameters $\beta_A$, $\beta_B$ and $\beta_C$.
Equation~(\ref{SVV}) is analogous to the $PVV$ interaction which was
originally introduced as a $\pi\rho\omega$ coupling a long time
ago~\cite{GSW}. It is intended to be the simplest 
 description of the production mechanism which contains the full
symmetries of the problem. Elsewhere we will discuss modifications due to
the effect of K-loops.
 One can now compute the amplitudes for
$S\rightarrow\gamma\gamma$ and $V \rightarrow S \gamma$ according to
the diagrams of Fig.~\ref{fig:diagrams}.

Altogether there are many processes of these types. For the two photon decays
one may consider the initial scalar to be any of $\sigma(560), f_0(980)$
or $a_0^0(980)$. With an initial vector state we have, in addition to
$\phi \rightarrow f_0,a_0^0 + \gamma$, the possibilities $\phi \rightarrow \sigma
+\gamma, \omega \rightarrow \sigma + \gamma$ and $\rho^0 \rightarrow \sigma + \gamma$.
Furthermore for the cases when the scalar may be heavier than the vector, the same
diagram allows one to compute the five modes $f_0, a_0^0 \rightarrow \omega,
\rho^0 + \gamma$ as well as $\kappa^0 \rightarrow K^{*0} +\gamma$. These are not
all measured yet but an initial predicted  correlation, is shown in \cite{BHS}.
 
  This model can also be used to study a recent
conjecture\cite{CK} which attempts to produce a large value for the ratio 
$\Gamma(\phi\rightarrow f_0\gamma)/\Gamma(\phi \rightarrow a_0\gamma)$
by invoking the iso spin violating $a_0(980)-f_0(980)$
mixing. Actually, a detailed refutation of this conjecture has already
been presented\cite{AK}. However the calculation may illustrate
our approach. One may simply introduce the mixing
by a term in the effective Lagrangian:
${ L}_{af} = A_{af} a_0^0 f_0$.
A recent calculation~\cite{ABFS} for the purpose of finding the effect
of the scalar mesons in the $\eta\rightarrow 3\pi$ process obtained
the value
$A_{af} = -4.66\times10^{-3}\,\mbox{GeV}^2$.
It is convenient to treat this term as a perturbation.
  Then the amplitude factor for $\phi\rightarrow f_0\gamma$ includes a correction term
consisting of the $\phi\rightarrow a_0^0\gamma$ amplitude factor
$C_\phi^{a_0} =
 \sqrt{2}\left(\beta_C - 2\beta_A\right)$ 
 multiplied by $A_{af}$ and by the $a_0$ propagator.
The $\phi\rightarrow a_0^0\gamma$ amplitude factor has a similar correction.
The desired ratio is then,
\begin{equation}
\frac{amp(\phi\rightarrow f_0\gamma)}{amp(\phi\rightarrow a_0^0\gamma)}=
\frac{ C_\phi^f + A_{af}C_\phi^a/D_a(m_f^2) }
{ C_\phi^a + A_{af}C_\phi^f/D_f(m_a^2)},
\label{amprat}
\end{equation}
where $D_a(m_f^2)=-m_f^2 + m_a^2 -im_a\Gamma_a$ and $D_f(m_a^2) =
-m_a^2 +m_f^2 -im_f\Gamma_f$. In this approach the propagators are diagonal in the
isospin basis.  
The numerical values of these
resonance widths and masses are,
 according to the Review of
Particle Physics~\cite{PDG} $m_{a_0} = (984.7\pm1.3)\,\mbox{MeV}$,
$\Gamma_{a_0} = 50$--$100$\,MeV, $m_{f_0} = 980 \pm10\,\mbox{MeV}$  
and $\Gamma_{f_0} = 40$--$100$\,MeV. For definiteness, from
 column 1 of Table II in Ref.~\cite{HSS} we
take $m_{f_0} = 987\,\mbox{MeV}$ and $\Gamma_{f_0} = 65\,\mbox{MeV}$   
while in Eq.~(4.2) of Ref.~\cite{FS}
we take $\Gamma_{a_0} = 70 \,\mbox{MeV}$.
In fact the main conclusion does not depend on these precise values.   
It is easy to see that the mixing factors are approximately  
given by
\begin{equation}
\frac{A_{af}}{D_a(m_f^2)} \approx \frac{A_{af}}{D_f(m_a^2)}   
\approx \frac{iA_{af}}{m_a\Gamma_a} \approx -0.07i.
\label{mfapprox}
\end{equation}
Noting that $C_\phi^f/C_\phi^a \approx 0.75 $ in the present model,
the ratio in Eq.(\ref{amprat}) is roughly $(0.75-0.07i)/(1-0.05i)$.
Clearly, the correction to $\Gamma(\phi \rightarrow f_0\gamma)/\Gamma(
\phi \rightarrow a_0\gamma)$ due to $a_0^0$-$f_0$ mixing only amounts  
to a  few per cent, nowhere near the huge effect suggested in\cite{CK}.

We are happy to thank N. N. Achasov for important communications
and A. Abdel-Rehim, A. H. Fariborz and F. Sannino for very
helpful discussions. Prof. Fariborz also is to be thanked for
excellently organizing this stimulating conference.
 D.B. wishes to acknowledge support from the Thomas Jefferson National
Accelerator Facility operated by the Southeastern Universities Research
Association (SURA) under DOE contract number DE-AC05-84ER40150.
The work of M.H.
is supported in part by Grant-in-Aid for Scientific Research
(A)\#12740144 and USDOE Grant Number DE-FG02-88ER40388. The
work of J.S. is supported in part by DOE contract DE-FG-02-85ER40231.

\end{document}